\input harvmac

\lref\BernUG{
  Z.~Bern, L.~J.~Dixon, D.~C.~Dunbar, M.~Perelstein and J.~S.~Rozowsky,
  ``On the relationship between Yang-Mills theory and gravity and its
  implication for ultraviolet divergences,''
  Nucl.\ Phys.\ B {\bf 530}, 401 (1998)
  [arXiv:hep-th/9802162].
}

\lref\BernZX{
  Z.~Bern, L.~J.~Dixon, D.~C.~Dunbar and D.~A.~Kosower,
  ``One loop $n$
  point gauge theory amplitudes, unitarity and collinear limits,''
  Nucl.\ Phys.\ B {\bf 425}, 217 (1994)
  [arXiv:hep-ph/9403226].
}

\lref\BernCG{
  Z.~Bern, L.~J.~Dixon, D.~C.~Dunbar and D.~A.~Kosower,
  ``Fusing gauge theory tree amplitudes into loop amplitudes,''
  Nucl.\ Phys.\ B {\bf 435}, 59 (1995)
  [arXiv:hep-ph/9409265].
}

\lref\BernDB{
  Z.~Bern and A.~G.~Morgan,
  ``Massive Loop Amplitudes from Unitarity,''
  Nucl.\ Phys.\ B {\bf 467}, 479 (1996)
  [arXiv:hep-ph/9511336].
}

\lref\BernFJ{
  Z.~Bern, L.~J.~Dixon and D.~A.~Kosower,
  ``Unitarity-based techniques for one-loop calculations in QCD,''
  Nucl.\ Phys.\ Proc.\ Suppl.\  {\bf 51C}, 243 (1996)
  [arXiv:hep-ph/9606378].
}

\lref\BernJE{
  Z.~Bern, L.~J.~Dixon and D.~A.~Kosower,
  ``Progress in one-loop QCD computations,''
  Ann.\ Rev.\ Nucl.\ Part.\ Sci.\  {\bf 46}, 109 (1996)
  [arXiv:hep-ph/9602280].
}

\lref\BernCZ{
  Z.~Bern, L.~J.~Dixon and D.~A.~Kosower,
  ``Two-loop $g \to g g$ splitting amplitudes in QCD,''
  JHEP {\bf 0408}, 012 (2004)
  [arXiv:hep-ph/0404293].
}

\lref\BuchbinderWP{
  E.~I.~Buchbinder and F.~Cachazo,
  ``Two-loop amplitudes of gluons and octa-cuts in ${\cal N} = 4$
  super Yang-Mills,''
  JHEP {\bf 0511}, 036 (2005)
  [arXiv:hep-th/0506126].
}

\lref\BernSC{
  Z.~Bern, V.~Del Duca and C.~R.~Schmidt,
  ``The infrared behavior of one-loop gluon amplitudes at
  next-to-next-to-leading order,''
  Phys.\ Lett.\ B {\bf 445}, 168 (1998)
  [arXiv:hep-ph/9810409].
}

\lref\KosowerXI{
  D.~A.~Kosower,
  ``All-order collinear behavior in gauge theories,''
  Nucl.\ Phys.\ B {\bf 552}, 319 (1999)
  [arXiv:hep-ph/9901201].
}

\lref\KosowerRX{
  D.~A.~Kosower and P.~Uwer,
  ``One-loop splitting amplitudes in gauge theory,''
  Nucl.\ Phys.\ B {\bf 563}, 477 (1999)
  [arXiv:hep-ph/9903515].
}

\lref\BernRY{
  Z.~Bern, V.~Del Duca, W.~B.~Kilgore and C.~R.~Schmidt,
  ``The infrared behavior of one-loop {QCD} amplitudes at
  next-to-next-to-leading order,''
  Phys.\ Rev.\ D {\bf 60}, 116001 (1999)
  [arXiv:hep-ph/9903516].
}

\lref\MagneaZB{
  L.~Magnea and G.~Sterman,
  ``Analytic Continuation Of The Sudakov Form-Factor In QCD,''
  Phys.\ Rev.\ D {\bf 42}, 4222 (1990).
}

\lref\BernNH{
  Z.~Bern, J.~S.~Rozowsky and B.~Yan,
  ``Two-loop four-gluon amplitudes in N = 4 super-Yang-Mills,''
  Phys.\ Lett.\ B {\bf 401}, 273 (1997)
  [arXiv:hep-ph/9702424].
}

\lref\MaldacenaRE{
  J.~M.~Maldacena,
  ``The large N limit of superconformal field theories and supergravity,''
  Adv.\ Theor.\ Math.\ Phys.\  {\bf 2}, 231 (1998)
  [Int.\ J.\ Theor.\ Phys.\  {\bf 38}, 1113 (1999)]
  [arXiv:hep-th/9711200].
}

\lref\CataniBH{
  S.~Catani,
  ``The singular behaviour of {QCD} amplitudes at two-loop order,''
  Phys.\ Lett.\ B {\bf 427}, 161 (1998)
  [arXiv:hep-ph/9802439].
}

\lref\SmirnovGC{
  V.~A.~Smirnov,
  ``Analytical result for dimensionally regularized massless on-shell double
  box,''
  Phys.\ Lett.\ B {\bf 460}, 397 (1999)
  [arXiv:hep-ph/9905323].
}

\lref\TauskVH{
  J.~B.~Tausk,
  ``Non-planar massless two-loop Feynman diagrams with four on-shell legs,''
  Phys.\ Lett.\ B {\bf 469}, 225 (1999)
  [arXiv:hep-ph/9909506].
}

\lref\AnastasiouKP{
  C.~Anastasiou, J.~B.~Tausk and M.~E.~Tejeda-Yeomans,
  ``The on-shell massless planar double box diagram with an irreducible
  numerator,''
  Nucl.\ Phys.\ Proc.\ Suppl.\  {\bf 89}, 262 (2000)
  [arXiv:hep-ph/0005328].
}

\lref\StermanQN{
   G.~Sterman and M.~E.~Tejeda-Yeomans,
   ``Multi-loop amplitudes and resummation,''
   Phys.\ Lett.\ B {\bf 552}, 48 (2003)
   [arXiv:hep-ph/0210130].
}

\lref\SmirnovKQ{
  V.~A.~Smirnov,
  ``Analytical evaluation of double boxes,''
  arXiv:hep-ph/0209177.
}

\lref\AnastasiouKJ{
  C.~Anastasiou, Z.~Bern, L.~J.~Dixon and D.~A.~Kosower,
  ``Planar amplitudes in maximally supersymmetric Yang-Mills theory,''
  Phys.\ Rev.\ Lett.\  {\bf 91}, 251602 (2003)
  [arXiv:hep-th/0309040].
}

\lref\WittenNN{
  E.~Witten,
  ``Perturbative gauge theory as a string theory in twistor space,''
  Commun.\ Math.\ Phys.\  {\bf 252}, 189 (2004)
  [arXiv:hep-th/0312171].
}

\lref\SmirnovBook{
  V.~A.~Smirnov,
  ``Evaluating Feynman Integrals,''
  Springer tracts in modern physics,
  {\bf 211} (Springer, Berlin, Heidelberg, 2004).
}

\lref\SmirnovIP{
  V.~A.~Smirnov,
  ``Evaluating multiloop Feynman integrals by Mellin-Barnes representation,''
  Nucl.\ Phys.\ Proc.\ Suppl.\  {\bf 135}, 252 (2004)
  [arXiv:hep-ph/0406052].
}

\lref\CachazoGA{
  F.~Cachazo and P.~Svrcek,
  ``Lectures on twistor strings and perturbative Yang-Mills theory,''
  PoS {\bf RTN2005}, 004 (2005)
  [arXiv:hep-th/0504194].
}

\lref\BernIZ{
  Z.~Bern, L.~J.~Dixon and V.~A.~Smirnov,
  ``Iteration of planar amplitudes in maximally supersymmetric Yang-Mills
  theory at three loops and beyond,''
  Phys.\ Rev.\ D {\bf 72}, 085001 (2005)
  [arXiv:hep-th/0505205].
}

\lref\AnastasiouCB{
  C.~Anastasiou and A.~Daleo,
  ``Numerical evaluation of loop integrals,''
  arXiv:hep-ph/0511176.
}

\lref\FriotCU{
  S.~Friot, D.~Greynat and E.~De Rafael,
  ``Asymptotics of Feynman diagrams and the Mellin-Barnes representation,''
  Phys.\ Lett.\ B {\bf 628}, 73 (2005)
  [arXiv:hep-ph/0505038].
}

\lref\CzakonRK{
  M.~Czakon,
  ``Automatized analytic continuation of Mellin-Barnes integrals,''
  arXiv: hep-ph/0511200.
}

\Title
{\vbox{
\baselineskip12pt
\hbox{hep-th/0601031}
\hbox{MCTP-05-106}
}}
{\vbox{
\centerline{Hidden Beauty in Multiloop Amplitudes}
}}

\centerline{
Freddy Cachazo${}^{1}$, Marcus Spradlin${}^{2,3}$
and Anastasia Volovich${}^{2}$
}

\bigskip

\centerline{
{\tt fcachazo@perimeterinstitute.ca},
~{\tt spradlin@ias.edu},
~{\tt nastja@ias.edu}
}

\vskip .5in
\centerline{${}^{1}$~Perimeter Institute for Theoretical Physics}
\centerline{Waterloo, Ontario N2L 2Y5, Canada}
\bigskip
\centerline{${}^{2}$~Institute for Advanced Study}
\centerline{Princeton, New Jersey 08540, USA}
\bigskip
\centerline{${}^{3}$~Michigan Center for Theoretical Physics}
\centerline{Ann Arbor, Michigan 48104, USA}

\vskip .5in
\centerline{\bf Abstract}

Planar $L$-loop maximally helicity violating
amplitudes in ${\cal N} = 4$ supersymmetric Yang-Mills
theory are believed to possess the remarkable property of
satisfying iteration relations in $L$.
We propose a simple new method for studying the iteration relations
for four-particle amplitudes which
involves the use of certain linear differential operators and
eliminates the need to fully evaluate any loop  integrals.
We carry out this procedure in explicit detail for the two-loop amplitude
and argue that this method can be used to prove the iteration relations
to all loops up to polynomials in logarithms.

\Date{January 2006}

\listtoc
\writetoc

\newsec{Introduction}

Maximally supersymmetric ${\cal N} = 4$ Yang-Mills (SYM) theory
possesses remarkably rich mathematical structure which
has been
the subject of intense investigation over the past several years.
One motivation for much of this work is the
AdS/CFT correspondence~\MaldacenaRE, which asserts that the strongly-coupled
SYM theory admits an equivalent formulation as 
gravity in AdS${}_5$,
thereby opening a new window
for studying quantum gravity. A complementary motivation, which has seen
a dramatic resurgence following \WittenNN\ (see \CachazoGA\ for a review),
is the desire to explore the mathematical
structure of Yang-Mills perturbation theory and to exploit that structure
to aid the calculation of scattering amplitudes.
Scattering amplitudes are the basic building blocks
which enter into the calculation of experimentally measured processes.
As the LHC comes on line in the next couple of years there will be increasing
pressure to bring the theoretical uncertainties in QCD calculations, especially
at higher loops, under control.

Optimistically, we hope that
the rich structure of Yang-Mills perturbation theory and the
simplicity of the strongly-coupled theory expected from AdS/CFT are
two sides of the same coin, and that we might in some cases be able
to see some hint of the structure
which enables perturbation theory to be resummed to match
onto AdS/CFT.

One intriguing step in this direction has been the study of
iterative relations amongst planar
maximally helicity violating (MHV)
loop amplitudes in dimensionally regulated
($d = 4 - 2 \epsilon$) ${\cal N} = 4$ SYM
\refs{\AnastasiouKJ,\BernIZ}.
In \AnastasiouKJ\ Anastasiou, Bern, Dixon and Kosower
suggested that two-loop MHV
amplitudes obey the iteration
\eqn\abdktwo{
M_n^{(2)}(\epsilon) = \ha \left( M_n^{(1)}(\epsilon) \right)^2 +
f^{(2)}(\epsilon) M_n^{(1)}(2 \epsilon) - {\pi^4 \over 72} +
{\cal O}(\epsilon),
}
where
we use $M_n^{(L)}(\epsilon) = A_n^{(L)}(\epsilon)/A_n^{(0)}$
to denote the ratio of the
$L$-loop $n$-particle
color-stripped partial
amplitude to the corresponding tree-level amplitude,
and
\eqn\fdef{
f^{(2)}(\epsilon) \equiv {1 \over \epsilon}( \psi(1 - \epsilon) - \psi(1))
= - (\zeta(2) + \zeta(3) \epsilon + \zeta(4) \epsilon^2 + \cdots ),
\qquad \psi(x) \equiv {d \over d x} \ln \Gamma(x).
}
We refer to \abdktwo\ for general $n$ as the ABDK conjecture.
In section 4 we review
an exponentiated ansatz conjectured in \BernIZ\ 
which generalizes the iterative structure \abdktwo\ for
MHV amplitudes to all loops.
These conjectures were motivated by studies of the infrared
\refs{\MagneaZB,\CataniBH,\StermanQN} and collinear
\refs{\AnastasiouKJ,\BernSC \KosowerXI  \KosowerRX-\BernRY}
behavior of multiloop amplitudes, where similar basic iterative structures
appear.

The problem of calculating $L>1$ amplitudes
and then
verifying these iterative structures involves
numerous extremely difficult technical challenges.
So far the only iteration relations which have been explicitly verified
are the two-loop ABDK relation \abdktwo\ for $n=4$ particles \AnastasiouKJ\ 
(the amplitude $M_4^{(2)}(\epsilon)$
was originally calculated in \BernNH\ using unitarity
based methods \refs{\BernZX \BernCG \BernDB \BernJE \BernFJ \BernUG-\BernCZ}),
and its three-loop generalization \BernIZ, again just for $n=4$, where
the calculation of $M_4^{(3)}(\epsilon)$
by Bern, Dixon and Smirnov
through ${\cal O}(\epsilon^0)$ is  an impressive tour de force.
Even at two loops the $n>4$ amplitudes are not yet known.
For some progress in this direction see \refs{\BernNH,\BuchbinderWP}.

In this paper we introduce new methods for studying these iteration
relations for $n=4$ particle amplitudes.
We  provide a direct check of the form of the ABDK
relation at two loops and then show that these methods can easily
be generalized to higher $L$, where they should greatly facilitate
the problem of verifying conjectured iteration relations.
We begin by expressing each amplitude appearing in \abdktwo\ in
a familiar integral form.  We then apply a linear differential
operator ${\cal L}^{(2)}$
which is chosen in such a way that
the terms in \abdktwo\ can be nicely combined into a 
single integral which is valid in a
neighborhood of $\epsilon = 0$.   The result \abdktwo\ then follows
by simply expanding in $\epsilon$ under the integral sign, eliminating
the need to fully evaluate any loop integrals.
Consequently, in our
calculation we never need to use the complicated
explicit formulas for 
the $\epsilon$ series expansions of
$M_4^{(1)}(\epsilon)$ and $M_4^{(2)}(\epsilon)$
in terms of polylogarithm functions.
The only ambiguity introduced
by acting with ${\cal L}^{(2)}$ is just a polynomial in
$\ln^2(-t/s)$,
which we show can be fixed up to an additive numerical
constant
by the known infrared and
collinear behavior of the two-loop amplitude.

In order to provide some motivation  and context for our calculation
we begin in section~2
with a more technical review
of the difficulties involved in checking \abdktwo\
and an outline of how
we propose to overcome them.
We present our calculation of the ABDK relation for $n=4$ in section 3.
Section 4 contains
a general argument which proves that
at any loop order $L$ it is always possible to construct a differential
operator ${\cal L}^{(L)}$ which generalizes the nice functionality of
${\cal L}^{(2)}$.  We
write down a particularly simple guess for a possible choice of
${\cal L}^{(L)}$ based on some very plausible 
assumptions about the structure of
higher loop amplitudes.

\newsec{Advanced Introduction and Motivation}

The ABDK conjecture is highly nontrivial because infrared divergences
in $M_n^{(1)}(\epsilon)$, which start at ${\cal O}(1/\epsilon^2)$, imply
that it is necessary to know $M_n^{(1)}(\epsilon)$ through
${\cal O}(\epsilon^2)$
in order to check \abdktwo\ 
through ${\cal O}(\epsilon^0)$.
The proof of \abdktwo\ for $n=4$ given in \AnastasiouKJ\ proceeds
by evaluating both
sides explicitly in terms of polylogs
and finding that they agree up to
terms of ${\cal O}(\epsilon)$.

Even for $n=4$,
the direct evaluation of the ABDK relation
is difficult for two reasons.
First of all,
both $M^{(2)}_4(\epsilon)$ and $M^{(1)}_4(\epsilon)$ are very complicated
functions of the kinematic ratio $x = -t/s$. Even using state-of-the-art
techniques, such as Mellin-Barnes representations,
the computation of the $\epsilon$ expansions
of these two amplitudes requires the
resummation of an infinite number of poles that give rise
to various polylogarithms.
To the order required to verify \abdktwo, these
amplitudes
involve polylogarithms of degree up to and including 4.
Secondly, even after $M^{(2)}_4(\epsilon)$ and $M^{(1)}_4(\epsilon)$ have both
been evaluated to the required order,
it is necessary to use nontrivial polylog identities to
check \abdktwo\ explicitly order by order.

The goal of this paper is to provide a far simpler method for
studying
iterative relations
which we hope may shed some light on the general
structure of these relations and may be useful to test
the conjectures in cases where their status is not currently known.
Our direct proof relies on  three basic ingredients.

First of all, we use the fact that
loop amplitudes
can be expressed in terms of Mellin-Barnes integral
representations (a detailed introductory treatment can be found in the book
\SmirnovBook).
These have the nice feature that the $x$
dependence can be isolated  in a simple manner.
In particular, any four-particle
loop amplitude can be written as an integral
of the form $\int_{-i\infty}^{+i \infty}
dy\ (-x)^y F(\epsilon,y)$ for some
function $F(\epsilon,y)$.
Often it is not very hard to find the function $F(\epsilon,y)$, rather it
is the final integral over $y$ which is exceedingly difficult to evaluate.
It is clearly tempting to
try to collect all of the terms in \abdktwo\ 
under a single $y$ integral
and check the ABDK conjecture
by expanding in $\epsilon$
`under the integral sign.'  In other words, we might like to look at
the {\it inverse} Mellin
transform\foot{Actually, since the contour for
$y$ is taken from $-i \infty$ to $+i \infty$, it is more appropriately
viewed as
a Fourier transform with respect to $\ln(-x)$ rather
than a Mellin transform with respect to $-x$.}
of
\abdktwo\ with respect to the variable $-x$.

However, this is usually not possible.
The reason is that Mellin-Barnes representations
of the form $\int_{-i\infty}^{+i \infty}
dy\ (-x)^y F(\epsilon,y)$ are
generally only valid for $\epsilon$ in an open set
which does not contain $\epsilon = 0$.
In order to make sense of statements such as \abdktwo, which involve
a series expansion around $\epsilon = 0$, it is necessary to analytically
continue the amplitudes in $\epsilon$.
Unfortunately,
as one analytically continues in $\epsilon$ towards $\epsilon = 0$
one frequently finds that along the way the $y$ integration contour will
hit poles in $F(\epsilon,y)$.  In crossing these
poles one
picks up residue terms which no longer have $y$ integrals.
In this paper we will refer to any such term which spoils our ability
to collect everything under a single $y$ integral and then expand
around $\epsilon = 0$  as an
{\it obstruction}.

We overcome this difficulty with a second ingredient:
linear differential operators which can be used to eliminate
all obstructions, or at least push them off so that they
are ${\cal O}(\epsilon)$ and can be ignored.
At two loops, the simplest example of an operator which accomplishes this
goal is $(x {d \over dx})^5$.
However we prefer to use the more elegant choice
\eqn\doperator{
{\cal L}^{(2)} \equiv
\left( x {d \over d x} - \epsilon \right)^3
\left( x {d \over d x} + \epsilon \right)^3
}
which has several attractive features, even though it is
of higher degree\foot{In fact there is even
a degree 4 operator $\left(x {d \over d x} - 
\epsilon \right)^2 \left(x {d \over d x} + \epsilon \right)^2$
which kills many, but not all, of the obstructions, leaving a remainder
which is just an innocent
number plus ${\cal O}(\epsilon)$.
This operator would be sufficient
for verifying the ABDK conjecture,
although it complicates the analysis slightly because one has to keep
track of the residues which are not completely killed but only injured.}.
Most importantly, we will see that ${\cal L}^{(2)}$ kills all
obstructions {\it exactly}, to all orders in $\epsilon$.
In fact we will see that the choice of ${\cal L}^{(2)}$ is completely natural
from the Mellin-Barnes integral representation
of the amplitudes:
acting with ${\cal L}^{(2)}$
explicitly removes a number of poles in $F(\epsilon,y)$.
In contrast, acting with $(x {d \over dx})^5$ 
does not kill any of the poles; the residues of all of the original poles
must still be calculated and shown to be ${\cal O}(\epsilon)$ before
one can be sure that they can be ignored.
Also  as a minor bonus
we note that ${\cal L}^{(2)}$ conveniently preserves the manifest
$x \to 1/x$ symmetry of four-particle amplitudes.

Armed with ${\cal L}^{(2)}$ we present an elementary proof of the identity
\eqn\abdkoper{{\cal L}^{(2)} \left[ M_4^{(2)}(\epsilon)
- \ha \left( M_4^{(1)}(\epsilon) \right)^2 \right] =
{\cal O}(\epsilon),  }
without needing to evaluate the $y$ integral.
The proof involves only the evaluation of a
finite (and small!) number of residues
and the use of a single Barnes lemma
integral\foot{Barnes lemmas are integrals of
products of $\Gamma$ functions that give rise to
more $\Gamma$ functions and their derivatives.}.
Clearly \abdkoper\ is a weaker statement than \abdktwo\ because we
can add to $M_4^{(2)}(\epsilon) - \ha (M_4^{(1)}(\epsilon))^2$ any function
$K(\epsilon,x)$ in the kernel of ${\cal L}^{(2)}$.
It is clear upon inspection that this kernel consists of just a simple
function of
$\ln^2(-x)$---therefore, all polylogarithms in the ABDK relation are
unambiguously fixed already by \abdkoper.
This involves nontrivial cancellation between polylogarithms appearing
in the two terms, but we promise the reader that not a single polylogarithm
will appear in this paper.  For us all of the cancellation in \abdkoper\ 
occurs under the $y$ integral.

The final ingredient in our proof
is the known
infrared and collinear behavior of the two-loop amplitude
$M_4^{(2)}$.  We will show that these constraints 
fix the ambiguity discussed above to be
of the form
$K = f^{(2)}(\epsilon) M_4^{(1)}(2 \epsilon) + C + {\cal O}(\epsilon)$,
where $C$ is
a numerical constant which our methods do not determine.
In this particular case we happen to know from the work
of~\AnastasiouKJ\ that $C = - \pi^4/72$, but in general
it would probably not be terribly difficult to determine $C$---once
it is known to be independent of $x$---by
performing the Mellin-Barnes integrals numerically
at any convenient value of $x$ (see for example
\refs{\AnastasiouCB,\CzakonRK}).

\newsec{The Two-Loop Calculation}

In this section we begin by proving the identity
\eqn\miniabdk{
{\cal L}^{(2)} \left[ M_4^{(2)}(\epsilon)
- \ha \left( M_4^{(1)}(\epsilon) \right)^2 \right] =
{\cal O}(\epsilon),
}
and then use the known infrared and collinear behavior of
$M_4^{(2)}$ to fix the ambiguity arising from the differential
operator ${\cal L}^{(2)}$.
Of course, one could verify this identity by directly
substituting the known expressions for $M_4^{(2)}(\epsilon)$ and
$\ha (M_4^{(1)}(\epsilon))^2$ through ${\cal O}(\epsilon)$ and hitting
them with ${\cal L}^{(2)}$.
Our goal in this section is to demonstrate that it is
straightforward to prove \miniabdk\ directly at the level of
the Mellin-Barnes
{\it integrand}, without having to first fully evaluate the Mellin-Barnes
integral in terms of polylogarithms.
Along the way we will see the special feature of the particular choice
of the differential operator ${\cal L}^{(2)}$ which makes this possible.

\subsec{The $\ha (M^{(1)}_4)^2$ Term}

Let us first look at the second term in \miniabdk.
Taking a convenient expression for the one-loop box integral from (7)
of \SmirnovKQ\ (see also
\refs{\SmirnovGC,\SmirnovIP}),
we find the following simple Mellin-Barnes representation for
the one-loop amplitude:
\eqn\mone{
\eqalign{
M_4^{(1)}(\epsilon)&=
-\ha {e^{\epsilon \gamma} \over  (s t)^{\epsilon/2}
 \Gamma(-2 \epsilon)}
\int {dz \over 2 \pi i}
\  (-x)^{z + \epsilon/2}
\Gamma(1 + \epsilon + z) \Gamma^2(z) \Gamma^2(-\epsilon-z) \Gamma(1-z).
}}
The contour for the $z$ integral runs from $-i \infty$ to $+i \infty$
and passes to the right of all poles of the two $\Gamma(\cdots + z)$ 
functions and to the left of all poles of the two $\Gamma(\cdots - z)$
functions.
The contour can be taken to be a straight
line parallel to the imaginary axis as long as the
arguments of all $\Gamma$ functions have a positive real part\foot{We
remind the reader that this requirement comes about in the following way.
Many of the $\Gamma$ functions appearing
in Mellin-Barnes integral representations originate from
integrals of the form $\int_0^1 d \alpha \ \alpha^{a-1} (1-\alpha)^{b-1}
= \Gamma(a)\Gamma(b)/\Gamma(a+b)$ over Feynman parameters $\alpha$.
This integral only converges for ${\rm Re}(a), {\rm Re}(b) > 0$.  Other
$\Gamma$ functions come directly from the Mellin-Barnes representation
of quantities such as $(X + Y)^{-\nu}$; the arguments of
the  $\Gamma$ functions arising in this way must also have
arguments with positive real
parts.}.
This is only possible if ${\rm Re}(\epsilon)< 0$, in which case
we can take $0 < {\rm Re}(z) <  - {\rm Re}(\epsilon)$.
Note that $M_4^{(1)}(\epsilon)$ is symmetric under $s \leftrightarrow t$,
or equivalently $x \leftrightarrow 1/x$, as can be seen by making
the change of variable $z \to -z - \epsilon$.

Using \mone\ we can write
\eqn\aaa{\eqalign{
{1 \over 2} \left( M_4^{(1)}(\epsilon) \right)^2 &=
{1 \over 8}
{e^{2 \epsilon \gamma} \over  (s t)^\epsilon \Gamma^2(-2 \epsilon)}
\int {dz \over 2 \pi i} {dw \over 2 \pi i} (-x)^{w - z}
\cr
&\qquad\qquad\times
\Gamma(1 + \epsilon + z) \Gamma^2(z) \Gamma^2(-\epsilon-z) \Gamma(1-z)
\cr
&\qquad\qquad\times
\Gamma(1 + \epsilon + w) \Gamma^2(w) \Gamma^2(-\epsilon-w) \Gamma(1-w).
}}
Now make the change of variables
\eqn\aaa{
z = u - \ha y, \qquad w = u + \ha y
}
to consolidate the $x$ dependence into the factor $(-x)^y$.
This is convenient because acting
with the differential operator ${\cal L}^{(2)}$
simply inserts a factor of $(y^2 - \epsilon^2)^3$ into the integral.
Note that we hold the product $s t$ fixed while differentiating with respect
to $x = -t/s$.
Therefore
\eqn\zzz{
{\cal L}^{(2)}\left[ {1 \over 2} \left( M_4^{(1)}(\epsilon) \right)^2\right] =
{1 \over 8}
{e^{2 \epsilon \gamma} \over  (s t)^\epsilon \Gamma^2(-2 \epsilon)}
\int {du \over 2 \pi i} {dy \over 2 \pi i} F(u, y)
}
where
\eqn\Fdef{
\eqalign{
F(u,y) &= (-x)^y (y^2 - \epsilon^2)^3
\Gamma(1 + \epsilon + u - \half y)
\Gamma(u - \half y)^2 \Gamma(-\epsilon-u + \half y)^2
\Gamma(1-u + \half y)
\cr
&\qquad \times
\Gamma(1 + \epsilon + u + \half y)
\Gamma(u + \half y)^2
\Gamma(-\epsilon-u - \half y)^2 \Gamma(1-u - \half y).
}}

Let us now review\foot{See \refs{\SmirnovGC,\TauskVH} for
pioneering work on how
this is implemented at two loops.
A very convenient program which automates these 
kinds of manipulations has recently been made
available by M.~Czakon
\CzakonRK.}
the procedure for manipulating Mellin-Barnes integrals
such as \zzz.
The formula \zzz\ defines a meromorphic function of $\epsilon$
for $\epsilon$ in an open set which does not contain $\epsilon = 0$.
As we take $\epsilon \to 0$, the integration contours become pinched
until at $\epsilon = 0$ they are forced to run through some poles
of the $\Gamma$ functions.  In order to
construct a formula for $(M_4^{(1)}(\epsilon))^2$
which is valid
in an open neighborhood of $\epsilon = 0$ we must analytically continue
\zzz\ in $\epsilon$.

For definiteness, let us choose ${\rm Re}(u - \ha y) > {\rm Re}(u + \ha y)>0$.
The contour ${\cal C}$ for the
$u$ and $y$ variables must be such that the argument of each $\Gamma$ function
has a positive real part.  This can be satisfied as long
as\foot{There is also a lower bound on ${\rm Re}(\epsilon)$.
We do not worry about this since we are always interested in pushing
$\epsilon$ towards 0, not away from it.
We can imagine starting out with 
${\rm Re}(\epsilon)$ infinitesimally  less than ${\rm Re}(-u-\ha y)$.}
${\rm Re}(\epsilon) < {\rm Re}(-u - \ha y) < 0$.
With this choice,
the first pole that hits one of the integration contours as we try
to take $\epsilon \to 0$ is the first pole of $\Gamma(- \epsilon - u
+ \ha y)$.
Passing the pole through this contour
produces two terms: the first term is the same integral as in \zzz, but
with a  contour ${\cal C}'$
that now passes to the {\it right} of the first pole
of $\Gamma(- \epsilon - u  + \ha y)$,  and the second term is the residue
of $F(u,y)$ at $u = - \epsilon + \ha y$.
The residue comes with a minus sign because the contour passes across
the pole from left to right, leaving a clockwise integral around the pole.
Therefore
we have
\eqn\www{
\int_{\cal C} {du \over 2 \pi i} {d y \over 2 \pi i} F(u,y)
= \int_{\cal C'} {du \over 2 \pi i} {d y \over 2 \pi i} F(u,y)
- \int_{\cal C} {dy \over 2 \pi i} {\rm Res}_{u = - \epsilon + \ha y}
F(u,y).
}
Now when ${\rm Re}(\epsilon)$
becomes larger than ${\rm Re}(-u + \ha y)$,
the contour ${\cal C'}$ can be taken to be the straight line ${\cal C}$.
Therefore we replace ${\cal C}'$ by ${\cal C}$ in \www\ to write
\eqn\fff{
\int_{\cal C} {du \over 2 \pi i} {d y \over 2 \pi i} F(u,y)
\Rightarrow \int_{\cal C} {du \over 2 \pi i} {d y \over 2 \pi i} F(u,y)
- \int_{\cal C} {dy \over 2 \pi i} {\rm Res}_{u = - \epsilon + \ha y}
F(u,y).
}
The notation $\Rightarrow$ is a reminder of the logic here: it
would be incorrect to write $=$, since that would require the
second term on the right-hand side to be zero.
The left- and right-hand sides of \fff\ represent the same
meromorphic function of $\epsilon$.  The left-hand side is valid
in a neighborhood of ${\rm Re}(\epsilon) < {\rm Re}(-u + \ha y)$ while
the right-hand side defines the analytic continuation of the left-hand
side
for a neighborhood ${\rm Re}(\epsilon) > {\rm Re}(-u + \ha y)$.

The only remaining pole which we encounter as we take $\epsilon$ to
zero is the first pole of
$\Gamma(-\epsilon
- u - \ha y)$, at $u = - \epsilon - \ha y$.
This gives
\eqn\hhh{\eqalign{
\int_{\cal C} {du \over 2 \pi i} {d y \over 2 \pi i} F(u,y)
&\Rightarrow \int_{\cal C} {du \over 2 \pi i} {d y \over 2 \pi i} F(u,y)
\cr
&- \int_{\cal C} {dy \over 2 \pi i} {\rm Res}_{u = - \epsilon - \ha y}
F(u,y)
- \int_{\cal C} {dy \over 2 \pi i} {\rm Res}_{u = - \epsilon + \ha y}
F(u,y),
}}
with the left-hand side defined for ${\rm Re}(\epsilon) < {\rm Re}(-u+\ha y)$
and the right-hand side defined for ${\rm Re}(\epsilon) > {\rm Re}(-u-\ha y)$.
There are no more contours
standing in the way of taking $\epsilon \to 0$, so 
the three terms on the right-hand side of \hhh\ can therefore be evaluated
in an open neighborhood of $\epsilon = 0$ by simply making a power
series expansion in $\epsilon$ under the $y$ integral.
The first term in \hhh\ (the double integral term) is manifestly
${\cal O}(1)$
because $F(u,y)$ itself is.  When we remember the prefactor
in front of the integral \zzz, which contains an explicit
$1/\Gamma(-2 \epsilon)$, we see that this term only contributes to \zzz\ at
${\cal O}(\epsilon)$, so we can ignore it.
All we have to do is evaluate the two residues on the second line of \hhh\ and
expand the result (including the prefactor in \zzz) through
${\cal O}(\epsilon^0)$, which gives
\eqn\uuu{\eqalign{
 (s t)^\epsilon {\cal L}^{(2)}\left[
 {1\over 2} \left( M_4^{(1)}(\epsilon) \right)^2\right]&=
 \int {dy\over 2 \pi i} \ (-x)^y
y^6 \Gamma(1 - y) \Gamma(-y)^2 \Gamma(y)^{*2} \Gamma(1 + y)
\cr
&\qquad \times
\left[
- {2 \over \epsilon} + y \left( f^{(2)}(y) - f^{(2)}(-y) \right) \right]
+ {\cal O}(\epsilon),
}}
where $f^{(2)}$ is the function
defined in \fdef.
It is very intriguing to see this function, which is related to the
infrared and collinear
behavior of two-loop amplitudes, emerge in an interesting
way from a one-loop amplitude squared.
We follow the notation of \SmirnovBook\ in using $\Gamma^*$ as a reminder
that the contour for the $y$ integral passes the first pole of
the indicated $\Gamma$ function $\Gamma^2(y)$ on the wrong side.
The appropriate contour for  \uuu\ is $-1 < {\rm Re}(y) < 0$,
reflecting a choice we made in setting up the original $u$ and $y$
contours in \zzz.

It may appear that the application of the differential operator
${\cal L}^{(2)}$, and the corresponding insertion
of the factor $(y^2 - \epsilon^2)^3$ into the integral,
was not important in this analysis.  In fact it plays a crucial role
in removing a third-order obstruction.
Without including
this factor in $F$ we would
have found
\eqn\aaa{
- {\rm Res}_{u = - \epsilon + \ha y} F(u,y) =
- {2 (-x)^\epsilon \Gamma(-\epsilon)^4 \Gamma(1 + \epsilon)^4
\over (\epsilon - y)^3}
+ {\cal O}( (\epsilon - y)^{-2}),
}
which contains an obstruction to taking $\epsilon \to 0$ in the form
of a triple pole at $y = \epsilon$.  A factor of $(y - \epsilon)^3$
is needed to kill this obstruction.  The
necessity of the other factor $(y + \epsilon)^3$
only becomes apparent 
in the analysis of $M_4^{(2)}(\epsilon)$, to which we now turn
our attention.

\subsec{The $M^{(2)}_4$ Term}

In the previous subsection we reviewed in detail the procedure for
manipulating Mellin-Barnes integrals.
In this section we will calculate ${\cal L}^{(2)} M^{(2)}(\epsilon)$
by the same procedure, although we will not show each step in as much
detail.
The full two-loop amplitude is given by the sum of the two terms
\BernNH
\eqn\fulltwoloop{
M_4^{(2)}(\epsilon) = {1 \over 4} s^2 t I_4^{(2)}(s,t) +
{1 \over 4} s t^2 I_4^{(2)}(t,s),
}
where $I^{(2)}_4(s,t)$ is the two-loop massless scalar box function.
We start with a convenient Mellin-Barnes representation for the
two-loop box function given by (11) of \AnastasiouKP\foot{Our formula
follows from theirs after the change of variables $\alpha = z_1 + z_2$,
$\beta = z_1 - z_2$, $\tau = z - z_1$.}:
\eqn\itwodef{
\eqalign{
& {1 \over 4} s^2 t I_4^{(2)}(s,t) =  {1 \over 2} {e^{2 \epsilon \gamma} \over
 (st)^{ \epsilon} \Gamma(-2 \epsilon)}
\int
{d \sigma \over 2 \pi i}
{d z \over 2 \pi i}
{d z_1 \over 2 \pi i}
{d z_2 \over 2 \pi i}
\  (-x)^{1 - \sigma + \epsilon}
\cr
&\qquad\times
{ \Gamma(-z_1-z_2) \Gamma(-z_1+z_2) \Gamma(2 + \epsilon +z+z_1) \over
\Gamma(1-z_1-z_2) \Gamma(1-z_1+z_2) }
{\Gamma(\sigma) \Gamma(1+\epsilon-z-z_1-\sigma) \over
\Gamma(1 - 2 \epsilon+z+z_1)}
\cr
&\qquad\times
\Gamma(-1-\epsilon-z-z_2) \Gamma(-1-\epsilon-z+z_2)
\Gamma(1+z-z_1) \Gamma(1+z+z_1)
\cr
&\qquad\times
\Gamma(-\epsilon+z+z_2+\sigma) \Gamma(-z+z_1-\sigma)
\Gamma(-\epsilon+z-z_2+\sigma) \Gamma(1-\sigma).
}}
The full two-loop amplitude \fulltwoloop\ is obtained by
adding together this quantity and the same expression with
$x$ replaced by $1/x$.
Our goal is to reduce this to a single integral over the variable
$\sigma$, since that is the only
variable which sets the $x$ dependence through
the factor $(-x)^{1 - \sigma + \epsilon}$.
The contour is fixed by requiring that the argument of each
function has a positive real part.
The final result does not depend
on the particular choice, but intermediate steps can look a little different.
For definiteness, we imagine taking the contour
${\rm Re}(z) = - 2/3$, ${\rm Re}(z_1) = - 1/8$, ${\rm Re}(z_2) = -1/16$
and ${\rm Re}(\sigma) = 1/2$.

A simple analysis of \itwodef\ shows that
all of the poles we hit as we analytically continue
$\epsilon$ to $0$ sit at only four different possible values of $z_2$,
\eqn\resone{
z_2 = 1 + \epsilon + z, \qquad
z_2 = - 1 - \epsilon - z, \qquad
z_2 = \epsilon - \sigma - z, \qquad
z_2 = -\epsilon + \sigma + z.
}
Our first step is therefore to deform the contour past these poles.
This leaves us with a sum of four residue terms, each of which is
only a triple integral, plus the original four-fold integral
\itwodef, which can now be evaluated in a neighborhood of $\epsilon = 0$.
In fact the four-fold integral is manifestly ${\cal O}(\epsilon)$ due to the
explicit $1/\Gamma(-2 \epsilon)$ factor sitting in front of \itwodef.
Therefore we only need to follow the four residues \resone.
Let's look at the first residue term,
\eqn\rrr{
\eqalign{
&{1 \over 4} s^2 t I_4^{(2)}(s,t)=
\ha {e^{2 \epsilon \gamma} \over (s t)^\epsilon
\Gamma(-2 \epsilon)}
\int {d\sigma \over 2 \pi i} {d z_1 \over 2 \pi i} {dz \over 2 \pi i}
\ (-x)^{1 - \sigma + \epsilon}
\cr
&\times
{ \Gamma(-1-\epsilon-z-z_1)
\Gamma(1+\epsilon+z-z_1)  \Gamma(2 + \epsilon+z+z_1)
\over
\Gamma(-\epsilon-z-z_1) \Gamma(2+\epsilon+z-z_1) }
{\Gamma(\sigma) \Gamma(1+\epsilon-z-z_1-\sigma) \over
\Gamma(1 - 2 \epsilon+z+z_1)}
\cr
&\times
\Gamma(-2-2 \epsilon - 2 z)
\Gamma(1+z-z_1) \Gamma(1+z+z_1)
\cr
&\times
\Gamma(1+2 z+\sigma) \Gamma(-z+z_1-\sigma)
\Gamma(-1-2 \epsilon+\sigma) \Gamma(1-\sigma)\cr
&\qquad
+ {\rm three~more~residues} + {\cal O}(\epsilon).
}}
This expression is now valid for $- 11/24 < {\rm Re}(\epsilon) < -1/3$
(given our specific choice for the contours).  As we continue taking
$\epsilon$ closer to zero we encounter more poles.  The ones we are
concerned about are poles which force us to take a residue in $\sigma$.
For example,
when ${\rm Re}(\epsilon)$ reaches $-1/4$ we 
hit a pole at $\sigma = 1 + 2 \epsilon$.
The residue at this pole is a double integral $\int dz_1 dz$.
The presence of this term
spoils our goal of trying to perform all manipulations
under the $\sigma$ integral.  

The dangerous
pole at $\sigma = 1+2 \epsilon$
comes from the $\Gamma(-1 - 2 \epsilon + \sigma)$
factor in \rrr.  We could get rid of this pole by inserting a factor
of $-1 - 2 \epsilon + \sigma$ into the integral.  But this is precisely
the factor we would get if we acted on \rrr\ with the differential
operator $-x {d\over dx} - \epsilon$!
Applying this differential 
operator kills the pole completely; the corresponding residue
is exactly zero, not just zero to ${\cal O}(\epsilon)$.

It is a straightforward exercise to continue following the Mellin-Barnes
procedure all the way to $\epsilon = 0$.
It turns out to be necessary to cube the factor
$(-1 - 2 \epsilon + \sigma)$ inside the integrand as this kills other
obstructions which appear at later steps.
This analysis motivates consideration of the differential operator
$(-x {d \over d x} - \epsilon)^3$.
However, we have only looked at the term ${1 \over 4} s^2 t I_4^{(2)}(s,t)$.
The complete two-loop amplitude \fulltwoloop\ 
is the sum of this and
${1 \over 4} s t^2 I_4^{(2)}(t,s)$, which
is the same up to $x \to 1/x$.  The same analysis applied to this
other term suggests that we should also hit $M_4^{(2)}(\epsilon)$
with the differential
operator $(x {d \over d x} - \epsilon)^3$.  Taken together, these
observations
motivate the choice of ${\cal L}^{(2)}$ in \doperator.

Let us now display the final
result for ${\cal L}^{(2)}$ acting on the Mellin-Barnes
integral \rrr.
Each of the triple integral terms in \rrr\ eventually branches
into the same triple integral, now valid in a neighborhood of
$\epsilon = 0$, plus a sum of residue terms which involve double  or
single integrals.
All four triple integral terms, and several
double integral terms which appear later on,
 turn out to be ${\cal O}(\epsilon)$, again
because of the explicit factor $1/\Gamma(-2 \epsilon)$ in \rrr.
However, this explicit factor $1/\Gamma(-2 \epsilon)$ certainly
does not mean that everything is ${\cal O}(\epsilon)$ since
taking residues can produce explicit singularities which cancel
this factor.
Ultimately we find that only 6 residues contribute through
${\cal O}(\epsilon)$,
\eqn\residues{\eqalign{
{\cal L}^{(2)} \left[ {1 \over 4} s^2 t I_4^{(2)}(s,t) \right] &=
{1 \over 2} {e^{2 \epsilon \gamma} \over (s t)^\epsilon \Gamma(-2 \epsilon)}
\Bigg[-
\int {d\sigma \over 2 \pi i}\ {\rm Res}_{z_1 = \epsilon}
{\rm Res}_{z = - 1 - \epsilon}
{\rm Res}_{z_2 = 1 + \epsilon + z}
\cr
&+
\int {d\sigma \over 2 \pi i}\ {\rm Res}_{z_1 = - \ha - \epsilon + \ha \sigma}
{\rm Res}_{z = - 1 - \epsilon - z_1}
{\rm Res}_{z_2 = 1 + \epsilon + z}\cr
& -
\int {d\sigma \over 2 \pi i}\ 
{\rm Res}_{z_1 = - \ha - \epsilon + \ha \sigma}
{\rm Res}_{z = - 1 - \epsilon - z_1}
{\rm Res}_{z_2 =-1 - \epsilon - z}\cr
&+
\int {d\sigma \over 2 \pi i}\ 
{\rm Res}_{z_1 = \epsilon}
{\rm Res}_{z = \epsilon  - \sigma}
{\rm Res}_{z_2 = \epsilon-z-\sigma}\cr
& +
\int {d\sigma \over 2 \pi i}
\int {d z_1 \over 2 \pi i}
\ 
{\rm Res}_{z = \epsilon  + z_1 - \sigma}
{\rm Res}_{z_2 = \epsilon-z-\sigma}\cr
&-
\int {d\sigma \over 2 \pi i}
\int {d z_1 \over 2 \pi i}\ 
{\rm Res}_{z = \epsilon + z_1 - \sigma}
{\rm Res}_{z_2 = -\epsilon+\sigma+z}
\Bigg]G
+ {\cal O}(\epsilon),
}}
where $G$ is the quantity appearing inside the integral in \itwodef.
The signs are easily fixed by keeping in mind that
a residue of the form ${\rm Res}_{v = \cdots \pm \epsilon}$ comes
with a sign of $\pm 1$.

A simple calculation reveals that the second and third line
in \residues\ 
are equal to each other, as are the fifth and sixth lines.
Let us look at these last two terms first since they still involve
an extra $z_1$ integral.  Expanding to ${\cal O}(\epsilon)$ we find
that they contribute
\eqn\yer{
\eqalign{
&{1 \over (s t)^\epsilon} \int 
{d\sigma \over 2 \pi i}
\ (-x)^{1 - \sigma}
 (1 - \sigma)^6
\Gamma(\sigma) \Gamma(\sigma-1) \Gamma^{*2}(1-\sigma) 
\cr
&\qquad \times
\int {dz_1 \over 2 \pi i}
\ 
\Gamma(2-\sigma + 2 z_1) \Gamma^*(2 z_1) \Gamma(-2 z_1) \Gamma^*(\sigma-1
- 2 z_1)
 + {\cal O}(\epsilon),
}}
where as before we keep track of the integration
contour by using a $*$ to denote
a $\Gamma$ function whose first pole is crossed in the `wrong' direction.
The remaining $z_1$ integral can be performed with the help of the generalized
Mellin-Barnes identity
\eqn\eee{
\eqalign{
&\int {dz \over 2 \pi i}\
\Gamma(\lambda_1 + z) \Gamma^*(\lambda_2 + z) \Gamma(-\lambda_2 - z)
\Gamma^*(\lambda_3 - z) \cr
&\qquad=
\Gamma(\lambda_2 + \lambda_3) \Big(
\Gamma(\lambda_1 - \lambda_2)
\left[ \psi(\lambda_1 - \lambda_2) - \psi(\lambda_1 + \lambda_3) \right]
- \Gamma(-\lambda_2-\lambda_3) \Gamma(\lambda_1 + \lambda_3)
\Big),
}}
which follows straightforwardly from (D.2) of \SmirnovBook.

Combining \yer\ with the first four terms in \residues, expanded
through ${\cal O}(\epsilon^0)$, gives
\eqn\ttt{
\eqalign{
& {\cal L}^{(2)} \left[ {1 \over 4} s^2 t I_4^{(2)}(s,t) \right]
=  {1 \over (s t)^\epsilon}
\int {d\sigma \over 2 \pi i}
\ (-x)^{1 - \sigma}
 (1 - \sigma)^6
\Gamma(\sigma) \Gamma^2(\sigma-1) \Gamma^{*2}(1-\sigma) \Gamma(2-\sigma)
\cr
&\qquad\qquad\qquad\qquad\qquad\times
\left[ -{1 \over \epsilon}
 + 4 \pi \cot(\pi(1- \sigma)) - \ln(-x) + \psi(2 - \sigma) - \psi(1)
\right]
+ {\cal O}(\epsilon).
}}

To get the full two-loop amplitude \fulltwoloop\ 
we should add \ttt\ to the same quantity with $x$ replaced by
$1/x$.  This looks equivalent to leaving $(-x)^{1-\sigma}$ alone
and replacing
$\sigma \to \sigma' = 2 - \sigma$ (and $\ln(-x) \to - \ln(-x)$)
inside the integral, but there is a subtlety.
The contour for $\sigma$ was chosen to run along ${\rm Re}(\sigma) = \ha$,
which would place the contour for $\sigma'$ at ${3 \over 2}$.
We would like
to rename $\sigma'$ back to $\sigma$ and combine both terms under
a single $\sigma$ integral.  In order to do this we must check that
we don't cross any poles in taking $\sigma'$ from ${3 \over 2} \to \ha$.
Fortunately the high power $(1 - \sigma)^6$ is more than adequate to kill
the pole at $\sigma = 1$, so there is no problem.

Finally we conclude that
\eqn\lmtwo{
\eqalign{
& {\cal L}^{(2)} \left[ M_4^{(2)}(\epsilon) \right]
=  {1 \over (s t)^\epsilon}
\int {d\sigma \over 2 \pi i}
\ (-x)^{1 - \sigma}
 (1 - \sigma)^6
\Gamma(\sigma) \Gamma^2(\sigma-1) \Gamma^{*2}(1-\sigma) \Gamma(2-\sigma)
\cr
&\qquad\qquad\qquad\qquad\qquad\times
\left[ -{2 \over \epsilon}
+  \psi(2 - \sigma) - 2 \psi(1) + \psi(\sigma)
\right]
+ {\cal O}(\epsilon).
}}
The change of variable $\sigma \to 1 - y$ 
and the identity
\eqn\aaa{
\psi(1-y) + \psi(1 + y) = \psi(y) + \psi(-y)
}
precisely
transform \lmtwo\ into \uuu.
This concludes the elementary proof that
\eqn\vvv{
{\cal L}^{(2)} \left[  M_4^{(2)}(\epsilon) - \ha \left(
M_4^{(1)}(\epsilon) \right)^2\right] = {\cal O}(\epsilon).
}
Note that while $M_4^{(2)}(\epsilon)$ and $\ha
(M_4^{(1)}(\epsilon))^2$ are both individually
${\cal O}(\epsilon^{-4})$, \uuu\ and \lmtwo\ indicate that
they become ${\cal O}(\epsilon^{-1})$ after being hit with
the differential operator ${\cal L}^{(2)}$.  This fact is evident
when one looks at the explicit formulas
\eqn\aaa{\eqalign{
(s t)^\epsilon
M_4^{(2)}(\epsilon) &= {2 \over \epsilon^4} - {1 \over \epsilon^2}
\left[ \ha \ln^2(-x) + {5 \over 4} \pi^2 \right] + {\cal O}(\epsilon^{-1}),
\cr
(s t)^\epsilon \ha \left(
M_4^{(1)}(\epsilon)\right)^2 &=
{2 \over \epsilon^4} - {1 \over \epsilon^2} \left[
\ha \ln^2(-x) + {4 \over 3} \pi^2 \right]
+ {\cal O}(\epsilon^{-1}).
}}

There is at least
one slightly intriguing aspect of the calculation in this
subsection.  In the calculation of
\AnastasiouKJ\ it was remarked that there is
a non-trivial cancellation of terms, involving the use of
polylogarithm  identities, between the two integrals $I_4^{(2)}(s,t)$
and $I_4^{(2)}(t,s)$ which contribute to the two-loop amplitude.
In our calculation this cancellation manifests itself 
 under the $\sigma$ integral  as
\eqn\aaa{
\cot(\pi(1-\sigma)) + \cot(\pi(\sigma - 1)) = 0.
}

\subsec{Fixing The Ambiguity}

The result \vvv\ implies that
\eqn\kilo{ M_4^{(2)}(\epsilon)
- \ha \left( M_4^{(1)}(\epsilon) \right)^2  ={1\over (st)^{\epsilon}} K(x,\epsilon) +
{\cal O}(\epsilon),}
where $K(x,\epsilon)$ is a conveniently
normalized undetermined function in the kernel of
${\cal L}^{(2)}$.  To conclude the check of the ABDK relation \abdktwo\ we
have to study this kernel.
Now
$K(x,\epsilon)$ has to be invariant under $x\to 1/x$ (since
the left-hand side of \kilo\ is), and it is easy to see that 
the most general function annihilated by ${\cal L}^{(2)}$ which respects
this symmetry is
\eqn\kernel{\eqalign{
K(x,\epsilon ) &= A(\epsilon )\left( (-x)^{\epsilon} +
(-x)^{-\epsilon} \right)\cr
&\qquad + B(\epsilon )\ln(-x) \left( (-x)^{\epsilon} -
(-x)^{-\epsilon} \right)\cr
&\qquad + C(\epsilon) \ln^2(-x) \left( (-x)^{\epsilon} +
(-x)^{-\epsilon} \right),
}}
where $A(\epsilon)$, $B(\epsilon)$ and $C(\epsilon)$ are
arbitrary functions of $\epsilon$ but
independent of $x$.
Note that \kernel\ is a function of $\ln^2(-x)$ only.

By inspecting the original relation \abdktwo\ it is clear that an important consistency
check of our method is to find that $(st)^{\epsilon}M^{(1)}_4(2\epsilon)$
can be expressed in the form
\kernel\ up to terms of order $\epsilon$. Indeed, a
little algebra shows that
\eqn\mone{(st)^{\epsilon} M^{(1)}_4{(2\epsilon)} = \left[-{1\over 4\epsilon^2}
+ {\pi^2\over 3}\right]\left( (-x)^{\epsilon} +
(-x)^{-\epsilon} \right) +{\ln(-x) \over 4 \epsilon} \left( (-x)^{\epsilon} -
(-x)^{-\epsilon} \right) +{\cal O}(\epsilon ). }
Equivalently, it is of course also possible to check directly that
${\cal L}^{(2)} M_4^{(1)}(2 \epsilon) = {\cal O}(\epsilon)$.
Since we have verified that
the quantity $(s t)^\epsilon M_4^{(1)}(2 \epsilon)$
lies in the kernel of ${\cal L}^{(2)}$ (to ${\cal O}(\epsilon)$), we can
freely absorb this into the ambiguity in $K(x,\epsilon)$ to rewrite
\kilo\ in the equivalent form
\eqn\kilotwo{ M_4^{(2)}(\epsilon)
- \ha \left( M_4^{(1)}(\epsilon) \right)^2 
- f^{(2)} M_4^{(1)}(2 \epsilon)
={1\over (st)^{\epsilon}} K(x,\epsilon) +
{\cal O}(\epsilon).}

Let us now take a look at what we can say about the remaining
ambiguity $K(x,\epsilon)$
from the infrared singularity structure of two-loop amplitudes.
This has been studied in QCD \refs{\CataniBH,\MagneaZB}
and can be applied to ${\cal N}=4$ SYM as well.
The known behavior
implies that
the quantity appearing on the left-hand side of \kilotwo\ is free of
IR divergences.
This implies that $K(x,\epsilon)$ must not have any poles in $\epsilon$.
Looking at the most general form \kernel\ of $K(x,\epsilon)$, it is easy
to see that 
the only function compatible with the infrared behavior is
\eqn\newre{ {1 \over (s t)^\epsilon}
K(x,\epsilon ) = C + E \ln^2(-x) + F \ln^4(-x) +  {\cal O}(\epsilon),}
where $C$, $E$  and $F$ are numerical constants.

In fact it is easy to reduce the ambiguity slightly \newre\ by
using a lower-order differential operator instead of ${\cal L}^{(2)}$.
For example, the  operator
$(x {d \over d x} + \epsilon)^2
(x {d \over d x} - \epsilon)^2$, which we mentioned in section 2, kills most,
but not all, of the obstructions, leaving as  remainders just numerical
constants (which can be absorbed into $C$ in \newre).
One could even consider just the operator $(x {d \over d x})^4$, which
does not completely kill any obstruction, but reduces all of them
to  numbers plus ${\cal O}(\epsilon)$.  The use of these operators would have
complicated the analysis of section 2, since we would have to keep track
of all of these unkilled obstructions.  Their only advantage is that
these operators have smaller kernels than ${\cal L}^{(2)}$, 
so one can use them to argue that $F$ in \newre\ must be zero.

An independent argument which reduces the ambiguity \newre\ even further
involves considering the
collinear limit of the two-loop amplitude $M_4^{(2)}(\epsilon)$.
The known behavior
\refs{\AnastasiouKJ,\BernSC \KosowerXI \KosowerRX-\BernRY}
implies that the quantity
on the left-hand side of \kilotwo\ must be finite as $x \to 0$ and
$x \to \infty$, which immediately fixes $E=F=0$.
In conclusion, we have verified
the form of
the ABDK relation
\eqn\last{ M_4^{(2)}(\epsilon) =
\ha \left( M_4^{(1)}(\epsilon) \right)^2
+ f^{(2)}(\epsilon)M^{(1)}_4{(2\epsilon)}
+ C + {\cal O}(\epsilon)}
up to an overall additive numerical constant.

\newsec{Higher Loops}

In this section we would like to tie together the threads which
have been weaving around throughout the analysis of sections 2 and 3
into a coherent picture.
We show  that the methods used in this paper
can be generalized
for $n=4$ particle amplitudes at
$L>2$ loops in a completely straightforward manner.
We first prove that
it is always possible
to find a generalization of the differential operator ${\cal L}^{(2)}$
which naturally removes
all obstructions to combining all of the terms in the
$L$-loop four-particle iteration relation inside a single integral.
In subsection 4.2 we will
write down a particularly simple form
for ${\cal L}^{(L)}$ which might be a natural guess to try.

\subsec{General Constructions}

The multiloop iteration relations proposed in \BernIZ\ take the form
\eqn\aaa{
M_n^{(L)}(\epsilon) = X_n^{(L)}[M_n^{(l)}(\epsilon)]
+ f^{(L)}(\epsilon) M_n^{(1)}(L \epsilon) + C^{(L)} + {\cal O}(\epsilon),
}
where $C^{(L)}$ are numerical constants, $f^{(L)}$ are functions
of $\epsilon$ only, and $X_n^{(L)}$ is a polynomial in the lower 
loop amplitudes which is conveniently summarized in the expression
\eqn\aaa{
X_n^{(L)}[M_n^{(l)}]
= M_n^{(L)} - \left.\ln \left( 1 + \sum_{l=1}^\infty
a^l M_n^{(l)} \right) \right|_{a^L~{\rm term}}.
}
For $L=2$ we recover \abdktwo\ while for $L=3$ we have for example
\eqn\threeloop{
M_n^{(3)}(\epsilon) = - {1 \over 3} \left( M_n^{(1)}(\epsilon)
\right)^3 + M_n^{(1)}(\epsilon) M_n^{(2)}(\epsilon)
+ f^{(3)}(\epsilon) M_n^{(1)}(3 \epsilon) + C^{(3)} +
{\cal O}(\epsilon).
}
This relation has been shown to be correct for $n=4$ by explicit
evaluation of the participating amplitudes \BernIZ.

We would like to collect all of the terms in \threeloop\ or its
higher $L$ generalizations, for $n=4$ particles,
under a single Mellin integral of the
form
\eqn\sss{
\int dy \ (-x)^y F(\epsilon,y)
}
and then to prove the iteration relations by
expanding in $\epsilon$ under the integral sign.
We found that in general this was not possible due to obstructions
which prevent us from taking $\epsilon \to 0$ inside the integral.
The most general possible obstruction is a residue of the form
\eqn\qpq{
{\rm Res}_{y = g(\epsilon)}[(-x)^y F(\epsilon,y)]
}
where $g(\epsilon)$ is determined by the precise arguments of the
$\Gamma$ functions appearing in $F(\epsilon,y)$.
In general it will be a linear function of $\epsilon$.
(For example, the obstruction we discussed below
\rrr\ corresponds to $g(\epsilon) = - \epsilon$ when the $\sigma$
variable is transcribed to the $y$ variable used in this section.)
The crucial feature of the expression \qpq\ is that while the resulting
$\epsilon$
dependence can be very complicated, the $x$ dependence is simple.
Let us suppose that the pole where we are taking the residue is a pole
of order $k$.  Then the residue \qpq\ will 
evaluate to
\eqn\obst{
(-x)^{g(\epsilon)} P_k(\ln(-x), \epsilon),
}
where $P_k$ is some polynomial in $\ln(-x)$ of degree $k$.  The coefficients
of that polynomial might have very complicated $\epsilon$ dependence.
A differential operator which exactly kills the residue \obst\ is 
\eqn\weirdop{
\left( x {d \over d x} - g(\epsilon) \right)^{k+1}.
}

In \qpq\ we analyzed the contribution from a single obstruction.
In a general amplitude (or product of amplitudes, such as appear in
the iteration relations) there will be many such obstructions.
Obviously it is always possible to construct a differential operator
which removes all of these obstructions by taking the product of
all of the individual differential operators of the form \weirdop---one
for each obstruction.

Now suppose that we would
like to test a conjectured iteration relation for $M_4^{(L)}(\epsilon)$ using
some differential operator ${\cal L}^{(L)}$ constructed in the manner
we have just described.
By construction, applying ${\cal L}^{(L)}$ to the iteration relation
will allow us to combine all of the terms under a single $y$ integral,
and then to verify the iteration relation by expanding the integrand
through ${\cal O}(\epsilon^0)$ inside the integral.
The ambiguity introduced by acting with ${\cal L}^{(L)}$ consists
precisely of terms of the form 
\obst, which span the kernel of ${\cal L}^{(L)}$ (by construction).
Since $g(\epsilon)$ is always a linear function of $\epsilon$, we
can expand \obst\ in $\epsilon$ to see that the only
things which can appear are powers of $\ln(-x)$ (actually,
the ambiguity must be a polynomial in $\ln^2(-x)$, due to the
$x \to 1/x$ symmetry).  There can never
be anything as complicated as a polylogarithm function---those
are uniquely fixed
even after ${\cal L}^{(L)}$ is applied.

In summary, we have demonstrated in this subsection that it is always
possible to construct a differential operator ${\cal L}^{(L)}$ which,
when applied to a conjectured iterative 
relation for $M_4^{(L)}(\epsilon)$, allows the relation to be checked without
explicitly evaluating the $y$ integral.
Moreover all of the complicated generalized polylogarithm functions are
completely fixed by this procedure; the only ambiguity
introduced by ${\cal L}^{(L)}$ is a polynomial in $\ln^2(-x)$.
It might be possible to argue away some of this ambiguity by
a more careful general argument.

\subsec{Specific Conjectures}

An example of an $L$-loop generalization
of ${\cal L}^{(2)}$ that would be natural to try is
\eqn\lloop{
{\cal L}^{(L)} = \left( x {d \over d x} +{L \over 2} \epsilon \right)^{L+1}
\left( x {d \over d x}- {L \over 2} \epsilon\right)^{L+1}
}
The form of this guess
is based on some observations of the particular structure
of the
Mellin-Barnes representations that one encounters at one, two, and
three loops.

The first observation is that obstructions tend to be of the form
\obst\ where in fact $g(\epsilon) = \pm L \epsilon/2$.  We are not
aware of any case where this is not true, so we conjecture that this
might be a general feature.  If true, it would mean that the
operator \lloop\ kills all obstructions {\it exactly}, to all orders
in $\epsilon$.
Even if \lloop\ does not kill all obstructions exactly,
it is clear that it is of sufficiently high order in $x { d \over dx}$
to push all obstructions to ${\cal O}(\epsilon^2)$ or higher,
where we do not care
about them.

\newsec{Summary and Outlook}

In this paper we have proposed to use certain simple linear differential
operators as effective tools for studying the iterative structure
of planar four-particle $L$-loop MHV amplitudes in ${\cal N} = 4$ SYM.
A key ingredient in our analysis is played by Mellin-Barnes integral
representations of loop amplitudes, and by what we call
`obstructions.'  Obstructions are terms which appear as we analytically
continue Mellin-Barnes integrals towards the region around $\epsilon = 0$;
their presence implies that it is impossible to collect everything
under a single integral which simultaneously preserves the $x$-dependence
in the simple factor $(-x)^y$ and also admits a series expansion in
$\epsilon$ under the $y$ integral.  We showed that obstructions are always
functions of $\ln^2(-x)$ and may be killed by the application of simple
differential operators.  The differential operator may be chosen to
be simply $x {d \over dx}$ to a sufficiently high power if one is
only interested in pushing the appearance of
obstructions to a sufficiently high
order in $\epsilon$, or a slightly more complicated operator if one
wishes to kill them exactly.  The operator ${\cal L}^{(2)}$ defined in
\doperator\ was proven to kill all obstructions appearing in the
two-loop ABDK iterative relation \abdktwo.
We argued that
the generalization
${\cal L}^{(L)}$ guessed in \lloop\ might achieve this task
at $L$ loops.

The advantage of removing obstructions is obvious:  by getting rid
of them, we can directly study the {\it inverse} Mellin transform
of an iterative relation in an expansion
around $\epsilon = 0$.  In other words we have no need to
evaluate the final, and often exceedingly complicated, $y$ integral
explicitly in terms of generalized polylogarithms.
The resulting simplicity is nicely exhibited by our elementary
verification of the structure of the two-loop ABDK relation \abdktwo\ for
$n=4$.
The proof of \vvv\ for 
required nothing more than evaluating a small
number of residues and using the single Mellin-Barnes integral \eee.

Fortunately, the price we pay for this beautiful simplification is not
too high: by construction, the only ambiguities introduced by the
differential operators we study are polynomials in $\ln^2(-x)$.
The complicated generalized polylogarithms are unambiguously fixed.
It is likely that the remaining $\ln^2(-x)$ ambiguity can in general
be fixed by the known infrared and collinear behavior of $L$-loop
amplitudes.

It is natural to suppose that these techniques
can also be generalized to $n > 4$ particle amplitudes.
In general these admit Mellin-Barnes representations where we isolate
the dependence on the various independent kinematic variables,
\eqn\aaa{
\int {d y_1 \over 2 \pi i} (k_1 \cdot k_2)^{y_1}
\int {d y_2 \over 2 \pi i} (k_1 \cdot k_3)^{y_2} \cdots
F(\epsilon; y_1,y_2,\cdots).
}
There will now be a more complicated picture involving complete obstructions
and partial obstructions in some subset of the variables, but it should
nevertheless be possible to construct partial differential operators
in several variables which might prove useful in studying the iterative
structure of these amplitudes.

Another interesting direction might be to use the converse mapping theorem to
compute the asymptotic behavior of Feynman integrals \FriotCU\ as
$t/s\to 0$ and as $t/s\to \infty$ in order to study possible
iterative relations in cases where proving the full structure is not within
reach\foot{We thank E.~De Rafael for bringing \FriotCU\ to our attention.}.
Particularly interesting would be to study the $n=4$ four-loop case.

\bigskip
\bigskip

\noindent
{\bf Acknowledgments}
\bigskip

FC thanks the IAS for hospitality and E.~Buchbinder for
extensive discussions on the ABDK conjecture.
MS and AV are grateful to R.~Roiban for
useful discussions on
Mellin-Barnes integrals.
The research of FC at the Perimeter Institute is supported in part
by funds from NSERC of Canada and MEDT of Ontario.
AV is supported by the William D.~Loughlin Membership at the IAS, and
MS and AV acknowledge support from the U.S.~Department of Energy under
grant number DE-FG02-90ER40542.

\listrefs
\end